# Dynamic Deformations and Forces in Soft Matter

Derek Y. C. Chan*[a,b], Evert Klaseboer[c] and Rogerio Manica[c]

[a] *Particulate Fluids Processing Centre, Department of Mathematics and Statistics, the University of Melbourne, Parkville, Victoria 3010, Australia. Fax: +61 3 8344 4599; Tel: +61 3 8344 5556; E-mail: D.Chan@unimelb.edu.au*

[b] *Department of Mathematics, National University of Singapore, 117543, Singapore. E-mail: D.Chan@unimelb.edu.au*

[c] *Institute of High Performance Computing, 1 Fusionopolis Way, 138632 Singapore. Fax: +65 6778 0522; Tel: +65 6419 1111; E-mail: evert@ihpc.a-star.edu.sg, manicar@ihpc.a-star.edu.sg*

**The ability of soft matter such as drops and bubbles to change shape dynamically during interaction can give rise to counter-intuitive behaviour that may be expected of rigid materials. Here we show that dimple formation on approach and the possibility of coalescence on separation of proximal drops in relative motion are examples of this general dynamic behaviour of soft matter that arise from the coupling between hydrodynamic forces and geometric deformations. The key parameter is a film capillary number $Ca_f \equiv (\mu V_o/\sigma)(R/H_o)^2$ that depends on viscosity $\mu$, interfacial tension $\sigma$, the Laplace radius $R$, characteristic film thickness $H_o$ and velocity $V_o$.**

Dynamic interactions involving deformable bodies such as drops, bubbles or elastic particles that are in close proximity to each other or to interfaces and boundaries is a central problem in soft matter physics. Such interactions involve time dependent flows and geometric deformations that occur simultaneously across very different length scales where nm thick films with μm lateral dimensions separate drops or particles ranging from 10's of μm to mm in size.

Recent studies of dynamic interactions include the use of: the surface force apparatus to measure the time-dependent profiles of thinning films between a deformable mercury drop in aqueous solution against a mica plate[1,2]; optical interference to visualize the dynamic stability of glycerol or water in silicone oil systems[3]; the atomic force microscope to measure dynamic forces between oil emulsion drops moving at typical Brownian speeds in aqueous electrolyte[4-6] and forces between bubbles and a solid substrate[7]; and the four-roll mill to manipulate interacting drops[8].

The above suite of experiments has been modelled numerically using the Stokes-Reynolds lubrication film theory plus the Young-Laplace equation for drop deformations. Key characteristics of thin films such as dimple formation[9-11], wimple excitation[12] and dynamic force measurements[4-7] have been predicted with good quantitative agreement. However, a recent drop coalescence study using a microfluidic cell revealed "a counter-intuitive phenomenon: coalescence occurs during the separation phase and not during the impact" and "there is no model that describes this phenomenon"[13].

In this communication we present a simple and physically perspicuous analysis of the Stokes-Reynolds Young-Laplace model that explains the general underlying physics of dynamic coalescence including the above counter-intuitive coalescence on separation as well as the onset of dimple deformations between approaching drops – a phenomenon that has been observed in experiments and numerical solutions. The key result can also be applied to deduce geometric deformations of interacting drops from measured forces.

Our approach is to use a perturbation analysis with matched asymptotic expansion of the Stokes-Reynolds Young-Laplace equations to derive a simple approximate solution which captures fully the essential physics and gives quantitative accuracy when compared to the full numerical solution of the governing equations. Consider two identical deformable drops separated by an axisymmetric film of the continuous phase. The radial, $r$ and time, $t$ evolution of the film thickness $h(r,t)$ and pressure $p(r,t)$ is, according to the Stokes-Reynolds lubrication model,

$$\frac{\partial h}{\partial t} = \frac{1}{12\mu r}\frac{\partial}{\partial r}\left(r\, h^3 \frac{\partial p}{\partial r}\right). \tag{1}$$

Implicit in Eqn. (1) is the assumption that the tangentially immobile ("no-slip") hydrodynamic boundary condition holds at the surface of the drops. This boundary condition has been shown to be consistent with a large number of experiments on the micro to nano scale[1-7,9,10,12]. Deformations of the drops are governed by the Young-Laplace equation in the inner film region where non-linear terms in the curvature may be omitted

$$\frac{\sigma}{2r}\frac{\partial}{\partial r}\left(r\frac{\partial h}{\partial r}\right) = \frac{2\sigma}{R} - p. \tag{2}$$

The Laplace pressure ($2\sigma/R$) defines the Laplace radius $R$.

We seek a formal solution of the form (see Inset of Fig. 1)

$$h(r,t) \equiv h_o(r,t) + h_1(r,t) \; ; \quad p(r,t) \equiv p_o(r,t) + p_1(r,t) \tag{3}$$

and by choosing $h_o(r,t) \equiv H(t) + r^2/R$ as the reference parabolic profile, Eqn. (2) becomes

$$\frac{\sigma}{2r}\frac{\partial}{\partial r}\left(r\frac{\partial h_1}{\partial r}\right) = -p. \tag{4}$$

Integration of Eqn. (4) gives the exact $r \to \infty$ asymptotic form: $h_1(r,t) \to -(F/\pi\sigma) \log(r)$, where $F = 2\pi \int_0^\infty rp \, dr$ is the hydrodynamic force between the two drops ($F > 0$, if repulsive). This logarithmic behaviour reflects the fact that Eqn. (2) is an inner equation for the film shape. The apparent divergence at large $r$ is to be matched to the outer solution that describes the drop shape outside the interaction zone of the film[14,15]. The appearance of the force $F$ in the pre-factor of the logarithm has been exploited recently to extract the total force exerted on a drop from the measured outer shape of its deformation[16].

The outer boundary condition required for the complete solution of Eqns. (1) and (2) or equivalently Eqns. (1) and (4), can be derived by imposing a constant volume constraint on the drop[4]. If one drop rests on a flat substrate where it subtends a contact angle $\theta$ and the substrate is moved relative to the other drop with a specified drive velocity $V(t)$ ($V > 0$ for separating drops), the outer boundary condition takes the form:

$$V(t) = \frac{\partial h(r_m,t)}{\partial t} + \frac{1}{2\pi\sigma}\left\{\log\left(\frac{r_m^2}{4R^2}\right) + 2B(\theta)\right\}\frac{dF}{dt} \quad (5)$$

where $r_m$ is some large radial position at the edge of the film and $B(\theta)$ is a known function of the contact angle $\theta$[17].

Theoretical predictions based on Eqns. (1), (2) and (5)[4-7,9,10,12,16] gave excellent agreement with different types of experimental studies of drop dynamics but the coupled partial differential equations had to be solved numerically.

We now derive a simple approximate analytical solution of Eqns. (1), (2) and (5) that provides informative physical insight into drop dynamics with quantitative precision.

The physical rationale in seeking a solution of the form of Eqn. (3) is to express the film thickness $h(r,t)$ formally as a non-deforming parabolic shape $h_o(r,t)$ whose location varies in time via $H(t)$. Deformations are described by $h_1(r,t)$ for which we derive a solution by perturbation. By setting $h(r,t) \approx h_o(r,t)$ in Eqn. (1) we find

$$p_o(r,t) = -\left(\frac{3\mu R}{2(H + r^2/R)^2}\right)\frac{dH}{dt}.$$

And using this result for $p$ in Eqn. (4) gives $h_1(r,t)$

$$h_1(r,t) = \frac{3\mu R^2}{4\sigma H}\frac{dH}{dt}\{\log(H + r^2/R) + C(t)\}. \quad (6a)$$

By integrating $p_o(r,t)$, the force is

$$F = 2\pi\int_0^\infty r\, p_o(r,t)\, dr = -\left(\frac{3\pi\mu R^2}{2H}\right)\frac{dH}{dt} \quad (6b)$$

so from Eqns. (6a) and (6b) we see that $h_1(r,t)$ has the expected logarithmic form: $h_1(r,t) \to -(F/\pi\sigma)\log(r)$, as $r \to \infty$.

The functions $H(t)$ and $C(t)$ can be determined by applying Eqn. (5) at the outer boundary $r_m$. Since $h_o(r,t)$ is a non-deforming parabolic shape at position $H(t)$, we have $\partial h_o(r,t)/\partial t = dH/dt = V(t)$ and $H(t) = H_o + \int_0^t V(\tau)\, d\tau$, where $H_o$ is the initial separation between the drops; and $C(t)$ can now be determined from Eqns. (5) and (6a). This then gives the desired solution:

$$h(r,t) \equiv h_o(r,t) + h_1(r,t) = [H(t) + r^2/R] + \left(\frac{3\mu R^2 V(t)}{4\sigma H(t)}\right)\{\log\left(\frac{H(t) + r^2/R}{4R}\right) + 2B(\theta)\} \quad (7)$$

This is the key result of this communication from which we can infer the characteristic behaviour of approaching and separating drops. As expected in matched expansion calculations, the result is independent of the precise value of matching position $r_m$, provided in this case, $r_m^2/R \gg H$.

We make two important observations regarding the film deformation $h_1(r,t)$ given by Eqn. (7):

(i) the magnitude of the film deformation $h_1(r,t)$ is characterized by the film capillary number: $Ca_f \equiv (\mu V_o/\sigma)(R/H_o)^2 \equiv Ca\ (R/H_o)^2$, where $V_o$ is a characteristic velocity;

(ii) the term in braces in Eqn. (7) is negative so time variations of the deformation $h_1(r,t)$ and the parabolic profile $h_o(r,t)$ have opposite signs whether the drops are separating ($V(t) > 0$ and $H(t)$ increasing) or are approaching ($V(t) < 0$ and $H(t)$ decreasing).

*Dimple formation on approach*. If the film capillary number $Ca_f$ is sufficiently large, the central portion of the film can thicken when the drops approach ($V(t) < 0$) and this is the physical origin for dimple formation between approaching drops. Numerical studies of drops approaching at constant velocity[10] identified a critical central film thickness $h_{dimple} \sim \alpha\ Ca^{1/2}R$ at which dimple formation will occur, where the numerical constant $\alpha$ is between $0.4 - 0.7$ for $Ca$ between $10^{-10}$–$10^{-4}$. While the perturbation solution Eqn. (7) predicts the same dependence on $Ca^{1/2}R$, the pre-factor is too large by an order of magnitude. This is not surprising since dimple formation actually occurs at separations where non-deforming drops would have overlapped.

*Coalescence on separation*. When the drops separate ($V(t) > 0$), the perturbation $h_1(r,t)$ will initially contribute a *decrease* to the central film thickness while $H(t)$ increases. For sufficiently large film capillary number $Ca_f$, the initial decrease in central film thickness can potentially bring the separation down to the range where the de-stabilizing influence of van der Waals attraction can take hold and initiate coalescence as observed in recent microfluidic cell experiments[13] or in four-roll mill experiments[8] where "coalescence frequently occurs during the part of the collision after the drops have already rotated to a configuration where they are being pulled apart by the external flow". Similar result has also been observed experimentally and verified theoretically for a mercury drop being separated from a mica surface[12]. Eqn. (7) therefore provides an approximate criterion as to when coalescence on separation can occur, namely when $Ca_f = (\mu V_o/\sigma)(R/H_o)^2 \sim 1$.

We now compare the predictions of Eqn. (7) for the evolution of the film with full numerical solutions of Eqns. (1), (2) and (5) using the following scales[4] to render these equations dimensionless and with all terms having the similar magnitude:

$$h \sim Ca^{1/2}R, \quad r \sim Ca^{1/4}R, \quad t \sim Ca^{-1/2}\mu R/\sigma, \quad p \sim \sigma/R . \tag{8}$$

We use a dimensionless velocity ramp $V(t) = v_o (1 - e^{-t/\tau})$ that accelerates the drops smoothly from rest to unit velocity: $v_o = 1$ (separation) or $-1$ (approach). The dimensionless position of the reference parabolic profile is $H(t) = H_o + v_o (t - 1 + e^{-t/\tau})$. We choose $\tau = 1$ (in dimensionless units), but since we are interested in times $t \gg 1$, the precise value of $\tau$ is not important. We choose the capillary number $Ca = 10^{-7}$, typical for experiments of dynamic deformations[1-3] and dynamic forces[4-7]. For simplicity, we set the contact angle $\theta = 90°$, so $B(\theta) = 1$[17].

In Fig. 1, we compare the total central film thickness $h(0,t) = h_o(0,t) + h_1(0,t)$ predicted by Eqn. (7) to the full numerical solution for two approaching drops ($v_o = -1$) from an initial dimensionless film thickness $H_o = 10$. Since for approaching drops $H(t)$ is decreasing, the analytic solution is only expected to hold well before the non-deforming parabolic profiles come into contact at $H(t) = 0$. In spite of this limitation, the perturbation solution of Eqn. (7) performs remarkably well down to a dimensionless thickness $h_o(0,t) \sim 5$ at the dimensionless time of 7. Beyond $t > 8$, the magnitude of the deformation is over predicted by the perturbation $h_1(0,t)$.

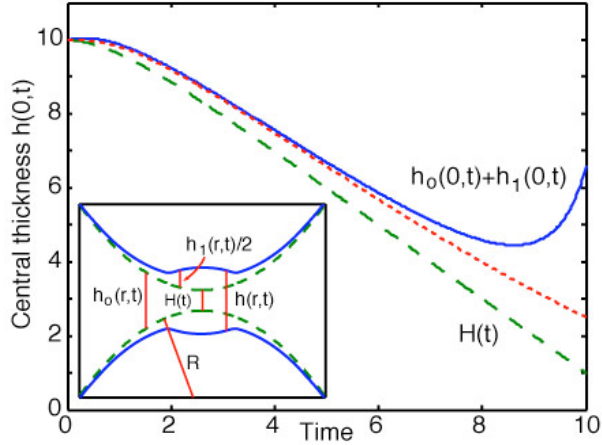

**Fig. 1** Approaching drops ($v_o = -1$): Time variations of the central film thickness, $h(0,t)$ of an axisymmetric film obtained from numerical solution (• •) of the film evolution equations (1), (2) and (5) compared to that of the reference parabolic shape: $h_o(0,t) = H(t) = H_o - (t - 1 + e^{-t/\tau})$ (– –) and the analytical solution: $h_o(0,t) + h_1(0,t)$) in Eqn. (7) (—). All quantities are dimensionless according to Eqn. (8).

A demonstration of the onset of *coalescence on separation* is given in Fig. 2 where we show variations of the dimensionless deformation $h_1(r,t)$ for separating drops ($v_o = 1$) from an initial film thickness $H_o = 10$. We see that the central deformation $h_1(0,t)$ becomes negative as the drops begin to separate, so deformations make the film *thinner* than that predicted by the parabolic profile $h_o(r,t)$. But as the separation progresses, $h_1(0,t)$ returns to zero after attaining a sharp minimum. If the magnitude of this minimum reduces the local film thickness sufficiently, coalescence can be initiated. There is good quantitative agreement between the analytical results in Eqn. (7) and the full numerical solution. Also the spatial form of $h_1(r,t)$ at various times marked on the $h_1(0,t)$ curve is reproduced rather accurately by Eqn. (7). This is perhaps not too surprising since the perturbation calculation is expected to be more accurate as the separation progresses.

In Fig. 3, we exhibit the onset of *dimple formation on approach* by showing variations of the dimensionless deformation $h_1(r,t)$ for approaching drops ($v_o = -1$) from an initial film thickness $H_o = 10$. In contrast to the case of separating drops, the central deformation $h_1(0,t)$ is positive and increases monotonically so that deformations make the film *thicker* than that predicted by the parabolic profile $h_o(r,t)$ as the drops approach. The prediction of the spatial form of $h_1(r,t)$ according to Eqn. (7) is satisfactory for times < 6. Note that for drops approaching at constant velocity, the reference parabolic profiles of the two drops will eventually come into contact and the perturbation method must then fail. For the case in Fig. 3, this occurs at dimensionless time ≈ 11.

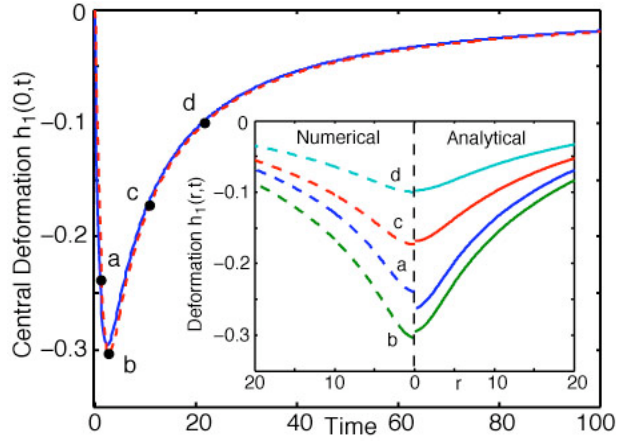

**Fig. 2**. Separating drops ($v_o = 1$): Time variations of the deformation, $h_1(0,t)$ at the centre of an axisymmetric film and (inset) a comparison of spatial variations of the deformation $h_1(r,t)$ at the indicated times (a to d) calculated by the analytical results in Eqn. (7) (—) and by a numerical solution (- -) of the film evolution equations (1), (2) and (5). All quantities are dimensionless according to Eqn. (8).

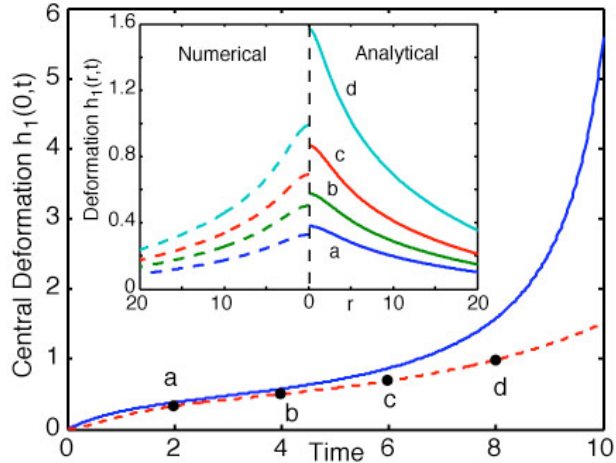

**Fig. 3** Approaching drops ($v_o = -1$): Time variations of the deformation, $h_1(0,t)$ at the centre of an axisymmetric film and (inset) a comparison of spatial variations of the deformation $h_1(r,t)$ at the indicated times (a to d) calculated by the analytical results in Eqn. (7) (—) and by a numerical solution (- -) of the film evolution equations (1), (2) and (5). All quantities are dimensionless according to Eqn. (8).

*Deformations from measured force.* If the force $F$ between two drops can be obtained, for example, from measurement with the atomic force microscope, the film profile $h(r,t)$ can be obtained by re-casting Eqns. (6b) and (7), with $V(t) = dH(t)/dt$, as

$$h(r,t) = [H(t) + r^2/R] - \frac{F}{2\pi\sigma}\{\log\left(\frac{H(t) + r^2/R}{4R}\right) + 2B(\theta)\} \qquad (9)$$

where $H(t)$ can be calculated from the way the drops are being driven in the experiment the experiment. To illustrate this idea, we show in Fig. 4 the film profile constructed with Eqn. (9) using the force obtained from our numerical calculation for the case shown in Fig. 3. This example illustrates the connection between the measured force $F$ and dynamic deformations of the film, and is complementary to an earlier approach whereby forces between deformed drops were calculated from drop geometry[16]. The utility of this approach is clear from the improvements in film profiles shown in Fig. 4 over those in Fig. 3 that was deduced from Eqn. (7). If, on the other hand, the drops are driven by a constant external force, $F_{ext}$ ($F_{ext} > 0$, to pull the drops apart), the function $H(t)$ can be found from solving: $F_{ext} = (3\pi\mu R^2/2H)(dH/dt)$ which gives: $H(t) = H_o \exp[(2F_{ext}/3\pi\mu R^2)t]$, a result that has been obtained earlier[14].

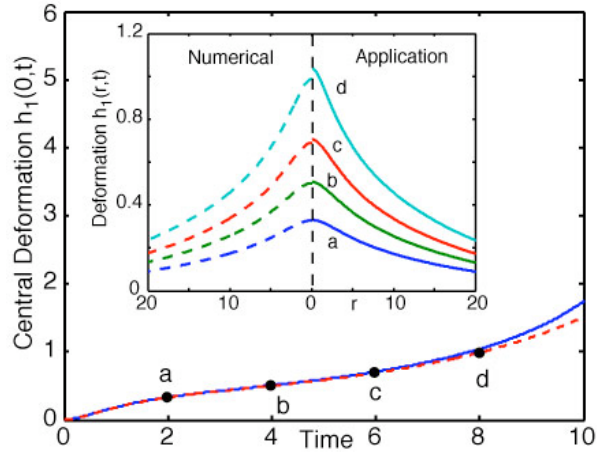

**Fig. 4** Approaching drops ($v_o = -1$): Time variations of central film thickness, $h(0,t)$ and (inset) the film profile, $h(r,t)$ of an axisymmetric film obtained from numerical solution (- -) of the film equations given by eqs. (1), (2) and (5) compared to that of the analytical solution (—) of Eqn. (9) using the numerically computed force. System parameters are the same as that in Fig. 3. All quantities are dimensionless according to Eqn. (8).

In this communication, we have derived a simple solution for the space-time evolution of the thin film between two drops as they approach or separate and elucidated the importance of coupling hydrodynamic interactions and geometric deformations that gave rise to the counter-intuitive phenomenon of coalescence on separation and the familiar dimple formation on approach. Although we have given results for the interaction between two identical drops, it is straightforward to generalize to the case of interacting dissimilar drops or to describe how drops interact with solids[17].

The above results are valid provided the film thickness is small compared to the characteristic radius of curvature so that the Stokes-Reynolds lubrication theory holds and the tangentially immobile ("no-slip") hydrodynamic boundary condition applies at the drop surface. As mentioned earlier, this boundary condition is consistent with a large number of experiments on the micro to nano scale[1-7,9,10,12]. For the case where the drop interface is mobile, internal flow in the drop will have to be taken into account by solving an integral equation relating the surface velocity to the surface stress[18,19]. In aqueous systems, inevitable surface impurities tend to arrest interfacial mobility and the mobile interface is not commonly encountered for drops or bubbles in the μm size range[20].

The concept of coupling between applied forces and geometric deformations giving rise to novel behaviour developed here can be generalized to soft matter bodies that deform because of elasticity or rearrangement of internal structures and interact via forces due to fluid flow, chemical or temperature gradients or due to applied magnetic or electrical fields together with differing material properties such as surface charge, magnetic susceptibility or dielectric permittivity.

Supported in part by Australian Research Council, AMIRA International and State Governments of Victoria and South Australia. DYCC is an Adjunct Professor at the National University of Singapore.